\documentclass[aps,preprint]{revtex4}%
\usepackage{amssymb}
\usepackage{amsfonts}
\usepackage{amsmath}
\usepackage{graphicx}%
\providecommand{\U}[1]{\protect\rule{.1in}{.1in}}

\begin{document}
\preprint{ }

\begin{center}
\bigskip

{\LARGE Can Quantum Nonlocality Be the Consequence of Faster-Than-Light
Interactions?}

{\large Luiz Carlos Ryff}

\textit{Instituto de F\'{\i}sica, Universidade Federal do Rio de Janeiro, Rio
de Janeiro , Brazil}

E-mail: ryff@if.ufrj.br

\bigskip

\bigskip

\textbf{Abstract}
\end{center}

{\small It has been advocated by Bell and Bohm that the
Einstein-Podolsky-Rosen (EPR) correlations are mediated through
faster-than-light (FTL) interactions. In a previous paper a way to avoid
causal paradoxes derived from this FTL hypothesis (via the breakdown of
Lorentz symmetry) has been suggested. Lorentz transformations would remain
valid, but there would be no equivalence between active and passive Lorentz
transformations in the case of EPR correlations. Some counterintuitive
consequences of this assumption are briefly examined here.}

\bigskip

In a previous paper \textrm{[1]} we investigated the idea advocated by Bell
and Bohm \textrm{[2]} according to which EPR correlations are mediated through
superluminal interactions. It has been shown that the formalism of quantum
mechanics leads to the conclusion that acting on one of the photons of an
entangled pair it is possible to force the other distant photon into a
well-defined polarization state. Although the argument is based on time-like
events, it seems reasonable to infer that such forcing does not cease to occur
in the case of space-like events, since the very same correlations are
observed. The consequence of assuming a finite speed for this FTL interaction
\textrm{[3]} has been critically analyzed, showing that the conclusion that it
leads to the possibility of superluminal communication is not inescapable.
Finally, a way to avoid causal paradoxes derived from the FTL hypothesis was
suggested via the breakdown of Lorentz symmetry. Lorentz transformations would
remain valid, but there would be no equivalence between active and passive
Lorentz transformations in the case of EPR correlations. I intend to examine
some consequences of this idea here.

As in \textrm{[1]}, we will consider a pair of reference frames, $\mathbf{S}$
and $\mathbf{S}^{\prime}$, in the standard configuration, where $\mathbf{S}$
is the privileged\textrm{\ }frame and $\mathbf{S}^{\prime}$ is the laboratory
frame moving with velocity\textrm{\ }$v<c$ along the\textrm{\ }$x$ axis, and
pairs of photons ($\nu_{1}$ and $\nu_{2}$), that propagate in opposite
directions, in the polarization-entangled state
\begin{equation}
\mid\psi\rangle=\frac{1}{\sqrt{2}}(\mid a_{\parallel}\rangle_{1}\mid
a_{\parallel}\rangle_{2}+\mid a_{\perp}\rangle_{1}\mid a_{\perp}\rangle_{2}),
\tag{1}%
\end{equation}
where $a_{\parallel}$ and $a_{\perp}$ represent arbitrary mutually orthogonal
directions. An interesting question is: Is it possible, being in
$\mathbf{S}^{\prime}$ and using EPR correlations, to determine $v$? The main
difficulty is that we cannot \textquotedblleft see,\textquotedblright\ so to
speak, when the second photon is forced (due to the action on the first
photon)\ into a well-defined polarization state. Furthemore, it is not
possible to know which photon is \textquotedblleft really\textquotedblright%
\ detected first \textrm{[4]}. I would like to examine here some curious and
counter-intuitive\ consequences of our basic assumption according to which the
FTL interaction propagates isotropically in $\mathbf{S}$ with a constant speed
$\overline{u}>c$, irrespective of the velocity of the source \textrm{[1]}. It
is instructive to see how things work.

\textbf{(A)} In the first situation to be considered, $\nu_{1}$ and $\nu_{2}$
are emitted at instant $t_{0}^{\prime}=0$\ from the source $S$, which is at
$x_{0}^{\prime}=0$ in $\mathbf{S}^{\prime}$, and propagate along the $x$ axis
in opposite directions. In $\mathbf{S}$ they are emitted at instant $t_{0}=0$
from $x_{0}=0$. Photon $\nu_{1}$ ($\nu_{2}$) is detected at point
$x_{1}^{\prime}=-l$ ($x_{2}^{\prime}=l$), with $l>0$, at instant
$t_{1}^{\prime}=l/c$ ($t_{2}^{\prime}=l/c$) \textrm{[5]}. The Lorentz
transformations connecting $\mathbf{S}$ and $\mathbf{S}^{\prime}$ are:%
\begin{equation}
x^{\prime}=\gamma\left(  x-vt\right)  , \tag{2}%
\end{equation}%
\begin{equation}
t^{\prime}=\gamma\left(  t-\frac{v}{c^{2}}x\right)  , \tag{3}%
\end{equation}%
\begin{equation}
x=\gamma\left(  x^{\prime}+vt^{\prime}\right)  , \tag{4}%
\end{equation}
and%
\begin{equation}
t=\gamma\left(  t^{\prime}+\frac{v}{c^{2}}x^{\prime}\right)  , \tag{5}%
\end{equation}
where $\gamma=1/\sqrt{1-v^{2}/c^{2}}$, from which we derive the expressions%
\begin{equation}
u_{x}^{\prime}=\frac{u_{x}-v}{1-\frac{vu_{x}}{c^{2}}} \tag{6}%
\end{equation}
and%
\begin{equation}
u_{x}=\frac{u_{x}^{\prime}+v}{1+\frac{vu_{x}^{\prime}}{c^{2}}} \tag{7}%
\end{equation}
for the velocities. Using $(4)$ and $(5)$ we see that in $\mathbf{S}$ the
photons are detected at
\begin{equation}
x_{1}=-\gamma(1-\frac{v}{c})l \tag{8}%
\end{equation}
and
\begin{subequations}
\begin{equation}
x_{2}=\gamma(1+\frac{v}{c})l \tag{9}%
\end{equation}
at instants
\end{subequations}
\begin{equation}
t_{1}=\gamma(1-\frac{v}{c})\frac{l}{c} \tag{10}%
\end{equation}
and
\begin{equation}
t_{2}=\gamma(1+\frac{v}{c})\frac{l}{c}. \tag{11}%
\end{equation}
Let us assume that $u_{x}=\overline{u}\rightarrow\infty$ for the velocity of
the FTL interaction in $\mathbf{S}$. Then, from $(6)$\ we obtain $\overline
{u}_{x}^{\prime}=-c^{2}/v$, for the velocity of the FTL interaction in
$\mathbf{S}^{\prime}$. Photon $\nu_{1}$ is detected at $t_{1}<$ $t_{2}$ in
$\mathbf{S}$. Whenever this takes place, $\nu_{2}$ is instantaneously forced
into a well-defined polarization state (since $\overline{u}\rightarrow\infty$
in $\mathbf{S}$), having traveled the distance $ct_{1}$. Using $(2)$ and
$(10)$ we see that in $\mathbf{S}^{\prime}$ it has traveled the distance
\begin{equation}
x_{F}^{\prime}=\gamma(ct_{1}-vt_{1})=\left(  \frac{c-v}{c+v}\right)  l.
\tag{12}%
\end{equation}
Therefore, in $\mathbf{S}^{\prime}$ photon $\nu_{2}$ \textquotedblleft
spontaneously\textquotedblright\ acquires a well-defined polarization state
when it is at $x_{F}^{\prime}$, at instant $t_{F}^{\prime}=x_{F}^{\prime}/c$
(in $\mathbf{S}^{\prime}$ $\nu_{1}$ has not yet reached the detection point)
\textrm{[6]}. Immediately an FTL interaction is triggered\ that goes from
$x_{F}^{\prime}$ to $x_{1}^{\prime}$, travelling the distance
\begin{equation}
x_{F}^{\prime}+\left\vert x_{1}^{\prime}\right\vert =\frac{2cl}{c+v} \tag{13}%
\end{equation}
in the time interval given by
\begin{equation}
t_{1}^{\prime}-t_{F}^{\prime}=\left(  \frac{2v}{c+v}\right)  \frac{l}{c}.
\tag{14}%
\end{equation}
Therefore, in $\mathbf{S}^{\prime}$ the FTL interaction propagates in the $-x$
direction with the speed
\begin{equation}
\frac{x_{F}^{\prime}+\left\vert x_{1}^{\prime}\right\vert }{t_{1}^{\prime
}-t_{F}^{\prime}}=\frac{c^{2}}{v}, \tag{15}%
\end{equation}
as it should be, and reaches $\nu_{1}$ exactly when it is being detected.

\textbf{(B) }In the second situation, we consider the same experiment
discussed in \textbf{(A)}, but now we are assuming $u_{x}=\overline{u}%
\neq\infty$, and $v$ is chosen to have $v\overline{u}/c^{2}=1$, which leads,
using $(6)$, to $\overline{u}_{x}^{\prime}\rightarrow\infty$.\textrm{\ }%
(\textit{It is worth noting that} $u_{x}=-\overline{u}$ \textit{leads to}
$\overline{u}_{x}^{\prime}=-(\overline{u}+v)/2$. \textit{The FTL interaction
does not propagate isotropically in} $\mathbf{S}^{\prime}$.) In $\mathbf{S}%
^{\prime}$, photons $\nu_{1}$ and $\nu_{2}$ are detected at the same time and
are then instantly connected by the FTL interaction which propagates with
infinite speed from $\nu_{1}(\nu_{2})$ to $\nu_{2}(\nu_{1})$. In $\mathbf{S}$,
photon $\nu_{1}$ is detected first, at instant $t_{1}$ given by $(10)$, which
triggers an FTL interaction sent in the direction of $\nu_{2}$. Let us
calculate the instant $t_{F}$\ when the interaction reaches the point at which
$\nu_{2}$ will be detected. The interaction is sent at instant $t_{1}$, it
then propagates to $x_{0}$ and then to $x_{2}$, given by $(9)$. From our
choice for $v$ we get $\overline{u}=c^{2}/v$, hence
\begin{equation}
t_{F}=\gamma\left(  1-\frac{v}{c}\right)  \frac{l}{c}+\gamma\left(  1-\frac
{v}{c}\right)  \frac{l}{\overline{u}}+\gamma\left(  1+\frac{v}{c}\right)
\frac{l}{\overline{u}}=\gamma\left(  1+\frac{v}{c}\right)  \frac{l}{c}=t_{2}.
\tag{16}%
\end{equation}
Therefore, in $\mathbf{S}$ the interaction reaches $\nu_{2}$ exactly when it
is being detected.

Apparently, there seems to be no practical way to distinguish between
situations \textbf{(A) }and \textbf{(B) }since, as previously observed, it is
not possible to know which photon is \textit{really} detected first, nor when
the FTL interaction reaches the second photon. Actually, to determine if
$\overline{u}\neq\infty$, the ideal is that in which the detection points are
equidistant from the source in the preferred frame. If photon $\nu_{1}$
($\nu_{2}$) is detected at point $x_{1}^{\prime}=-l_{1}$ ($x_{2}^{\prime
}=l_{2}$), with $l_{1}$ ($l_{2}$)$>0$, replacing $l$ by $l_{1}$ ($l_{2}$)\ in
$(8)$ ($(9)$) and making $-x_{1}=x_{2}$ we obtain
\begin{equation}
l_{1}=\left(  \frac{c+v}{c-v}\right)  l_{2}, \tag{17}%
\end{equation}
as the best choice.

It is interesting to reexamine situation \textbf{(B)}. If $\overline{u}$ is
finite and the detection points are equidistant from the source in the
preferred frame, no EPR correlations are to be expected. From $(17)$ we see
that in the laboratory frame $\nu_{2}$ is detected first, which triggers an
FTL interaction in the direction of $\nu_{1}$. As already emphasized, this
interaction propagates with a finite speed equal to $(\overline{u}+v)/2$, and
it is easy to verify that it\ cannot reach $\nu_{1}$ before it is detected. On
the other hand, when $\nu_{1}$ is detected, an FTL interaction with infinite
speed is sent, but it cannot reach $\nu_{2}$ since it has already been
detected. Therefore, although we have two different interpretations for the
same experiment, depending on the reference frame we use to describe it, they
lead to the same predictions. One possible difficulty is that the FTL speed
can be exceedingly large and, strictly speaking, the photons are never
detected at exactly the same time. Therefore, even observing EPR correlations,
it is not possible to conclude with absolute certainty that $\overline
{u}\rightarrow\infty$. On the other hand, if no EPR correlations are observed,
the next step would be to change the relationship between $l_{1}$ and $l_{2}$
to make the correlations appear.

\textbf{(C) }In the third situation, we will consider that the detection
points are along the $y$ axis in $\mathbf{S}^{\prime}$ ($y_{1}^{\prime}%
=-l_{1}$, $y_{2}^{\prime}=l_{2}$) and the source of the entangled photons is
at $y_{0}^{\prime}=0$. In the standard configuration, we have
\begin{equation}
y^{\prime}=y. \tag{18}%
\end{equation}
Hence, using $(18)$ and $(5)$ we obtain
\begin{equation}
u_{y}=\frac{u_{y}^{\prime}/\gamma}{1+\frac{vu_{x}^{\prime}}{c^{2}}}. \tag{19}%
\end{equation}
Assuming $u_{x}^{\prime}=\overline{u}_{x}^{\prime}=0$, and $u_{y}^{\prime
}=\overline{u}_{y}^{\prime}$, from $(7)$ and $(19)$ we obtain
\begin{equation}
\overline{u}_{x}=v \tag{20}%
\end{equation}
and
\begin{equation}
\overline{u}_{y}=\overline{u}_{y}^{\prime}\left(  1-\frac{v_{2}}{c_{2}%
}\right)  ^{1/2}. \tag{21}%
\end{equation}
Since $\overline{u}_{x}^{2}+\overline{u}_{y}^{2}=\overline{u}^{2}$, from
$(20)$ and $(21)$ we obtain
\begin{equation}
\overline{u}=\left[  v^{2}+\left(  \overline{u}_{y}^{\prime}\right)
^{2}\left(  1-\frac{v^{2}}{c^{2}}\right)  \right]  ^{1/2}. \tag{22}%
\end{equation}
Hence, knowing $\overline{u}_{y}^{\prime}$,\ the speed of the FTL interaction
in $\mathbf{S}^{\prime}$ along the $y$ axis, and $v$, the speed of the
laboratory frame relative to $\mathbf{S}$, the preferred frame, it would be
possible to determine $\overline{u}$, namely the speed of the FTL interaction
in $\mathbf{S}$.

\begin{center}%
\[%
{\includegraphics[
height=1.3889in,
width=2.597in
]%
{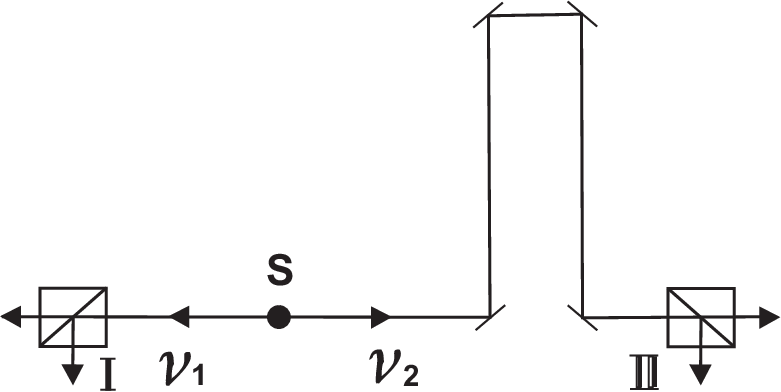}%
}
\]
Fig.1
\end{center}

Returning to the question posed at the beginning of this article, it seems
possible, at least in principle, to determine the velocity of the moving frame
$\mathbf{S}^{\prime}$ relative to the privileged frame $\mathbf{S}$, provided
that the speed $\overline{u}$ of the superluminal interaction in $\mathbf{S}$
is finite [7]. It is possible to devise a way to try to determine
$\overline{u}^{\prime}$, the speed of the superluminal interaction in
$\mathbf{S}^{\prime}$ (not forgetting that, contrary to what occurs in
$\mathbf{S}$, it will depend on the direction of propagation in $\mathbf{S}%
^{\prime}$). I believe it has become evident that this is a more complicated
task than it may seem at first sight. With this in mind, we can imagine the
following experiment. Let us consider a modified version of the experiment
depicted in Fig.1 [A source (S) emits a pair of polarization-entangled photons
($\nu_{1}$ and $\nu_{2}$) that propagate in opposite directions and impinge
respectively on two-channel polarizers (I and II). A detour is introduced to
have time-like events in which $\nu_{1}$\ is always detected before $\nu_{2}%
$.]. A second detour is introduced between the source and polarizer I. The
height of the detours can be adjusted continuously. Initially (Fig.1), the
height of the left detour is zero, and the height of the right\textrm{\ }%
detour is chosen to have $\nu_{1}$ being detected before $\nu_{2}$ in all
Lorentz frames (time-like events). We assume that in the first situation the
supposed superluminal interaction propagates from left to right. We then go on
increasing the height of the left detour continuously while, at the same time,
we decrease the height of the right detour continuously. Continuing with this
process, we will arrive at a situation in which it is now $\nu_{2}$ that is
detected before $\nu_{1}$ in all Lorentz frames. In this second situation the
superluminal interaction propagates from right to left. It is to be assumed
that between situations one and two there must be a region (involving
space-like events) in which the superluminal interaction no longer has an
effect. (Interestingly, in this experiment the detectors do not need to be far
from the source.) In principle, this would allow us to determine $\overline
{u}^{\prime}$. Naturally, since there is no isotropy, we would have\textrm{\ }%
different values for $\overline{u}^{\prime}$ propagating from left to right
and from right to left. In addition, rotating the experimental apparatus, we
would obtain other values for $\overline{u}^{\prime}$. To see this, instead of
using $(2)$ and $(3)$, we can use the equations below, that connect
$\mathbf{S}$ to $\mathbf{S}^{\prime}$ \textquotedblleft for the general case
where the $x$-axis is not in the direction of the velocity $v$%
\textquotedblright\ [8],
\begin{equation}
\mathbf{r}^{\prime}=\mathbf{r+}\frac{1}{v^{2}}(\gamma-1)(\mathbf{r.v}%
)\mathbf{v}-\gamma\mathbf{v}t \tag{23}%
\end{equation}
and
\begin{equation}
t^{\prime}=\gamma(t-\frac{\mathbf{r.v}}{c^{2}}), \tag{24}%
\end{equation}
which leads to
\begin{equation}
\mathbf{u}^{\prime}=\frac{1}{\gamma(1-\mathbf{u.v/}c^{2})}\left[
\mathbf{u}+\frac{1}{v^{2}}(\gamma-1)(\mathbf{u.v})\mathbf{v}-\gamma
\mathbf{v}\right]  . \tag{25}%
\end{equation}
From the standpoint of $\mathbf{S}$, if $\mathbf{u}=\overline{\mathbf{u}}$, we
obtain $\overline{\mathbf{u}}.\mathbf{v}=\overline{u}v\cos\theta$, where
$\theta$ is the angle between the direction of propagation of the superluminal
interaction and the direction of propagation of $\mathbf{S}^{\prime}$. In
short, no contradiction seems to arise from the breaking of\textrm{\ }%
equivalence between active and passive Lorentz transformations in the case of
EPR correlations. We merely have to keep in mind that the \textquotedblleft
correct\textquotedblright\ explanation, so to speak, is the one based on what
occurs in the privileged frame of reference. In principle, it is possible to
determine $\overline{u}$, namely the speed of the superluminal interaction in
$\mathbf{S}$, and $v$, the speed \ of $\mathbf{S}^{\prime}$ relative to
$\mathbf{S}$, using the equation that connects the velocities in
$\mathbf{S}^{\prime}$ to the velocities in $\mathbf{S}$:%
\begin{equation}
\mathbf{u}=\frac{1}{\gamma(1+\mathbf{u}^{\prime}\mathbf{.v/}c^{2})}\left[
\mathbf{u}^{\prime}+\frac{1}{v^{2}}(\gamma-1)(\mathbf{u}^{\prime}%
\mathbf{.v})\mathbf{v}+\gamma\mathbf{v}\right]  . \tag{26}%
\end{equation}
Since $\overline{\mathbf{u}}.\overline{\mathbf{u}}=\overline{u}^{2}%
=const.[\overline{u}\neq\overline{u}(\theta)]$, measuring $\overline
{u}^{\prime}$ for n different directions we obtain n equations which, at least
in principle, would allow us to determine $\mathbf{v}$\textbf{.} Since
$\mathbf{v.v}=v^{2}$ and\ $\overline{\mathbf{u}}^{\prime}.\mathbf{v}%
=\overline{u}^{\prime}v\cos\theta^{\prime}$, the unknowns are $v$ and
$\theta^{\prime}$, where $\theta^{\prime}$ is the angle between the direction
of propagation of the superluminal interaction, seen from $\mathbf{S}^{\prime
}$, and the direction of propagation of $\mathbf{S}^{\prime}$. In the first
measurement we have an unknown $\theta^{\prime}$, in the second we can choose
$\theta^{\prime}+\pi$ (rotating the apparatus), in the third, $\theta^{\prime
}+\pi/2$ (performing a new rotation), for instance, and so on. Strictly
speaking, we have more equations than unknowns. Actually, for each orientation
of the experimental apparatus we have a different equation. But the
experimental evidence we have so far seems to indicate that the quantum
entanglement holds for arbitrary distances, which strongly suggests that
entangled particles constitute a single entity ($\overline{u}\rightarrow
\infty$).

\bigskip

\begin{center}
\textbf{References\bigskip}
\end{center}

[1] L. C. Ryff, Einstein-Podolsky-Rosen (EPR) Correlations and Superluminal
Interactions, arXiv:1506.07383 [quant-ph].

[2] Interviews with John Bell and David Bohm in The Ghost in the Atom, P. C.
W. Davies and J. R. Brown (eds.), Cambridge University Press, Cambridge (1989).

[3] V. Scarani, N. Gisin, Phys. Lett. A 295, 167 (2002); V. Scarani, N. Gisin,
Braz. J. Phys. 35, 328 (2005); J.-D. Bancal, et al., Nat. Phys. 8, 867 (2012);
T. J. Barnea, et al.: Phys. Rev. A 88, 022123 (2013).

[4] Naturally, we may assume, tentatively, that $v$ = $V_{CMB}$, where the
acronym CMB refers to the Cosmic Microwave Background radiation, and $V_{CMB}
$ is the velocity relative to the frame (supposedly $\mathbf{S}$) in which
this radiation propagates isotropically. This would allow us to determine
which photon is \textquotedblleft really\textquotedblright\ detected first
(that is, which photon is first detected in $\mathbf{S}$).

[5] For the sake of simplicity, we are assuming that the FTL interaction is
triggered when the first photon is detected. But, strictly speaking, whether
the triggering occurs at the polarizer or at the detector can be considered an
open question.

[6] This suggests the possibility that an apparent random phenomenon in the
laboratory frame may actually be the consequence of a deterministic process in
the preferred frame.

[7] Strictly speaking, there can be no infinite speed, since infinite is a
limit, not an actual value. If entangled particles are instantly connected
this means that, somehow, they are a single entity, even if they are
arbitrarily distant from each other.

[8] Pauli, W.: Theory of Relativity. Pergamon Press (1967).

\end{document}